\documentclass[twocolumn]{pasj00}
\usepackage[dvips]{color}

\begin{document}
\SetRunningHead{N. Narita et al.}{Spin-Orbit Alignment of
the TrES-4b}
\Received{2010/02/04}
\Accepted{2010/03/11}

\title{
Spin-Orbit Alignment of the TrES-4 Transiting Planetary System\\
and Possible Additional Radial Velocity Variation$^*$}


\author{
Norio \textsc{Narita},\altaffilmark{1,2}
Bun'ei \textsc{Sato},\altaffilmark{3}
Teruyuki \textsc{Hirano},\altaffilmark{4}
Joshua N.\ \textsc{Winn},\altaffilmark{5}\\
Wako \textsc{Aoki},\altaffilmark{1} and
Motohide \textsc{Tamura},\altaffilmark{1}
}

\altaffiltext{1}{
National Astronomical Observatory of Japan, 2-21-1 Osawa,
Mitaka, Tokyo, 181-8588, Japan
}

\altaffiltext{2}{
Kavli Institute for Theoretical Physics, UCSB, Santa Barbara,
CA 93106-4030, USA
}

\altaffiltext{3}{
Global Edge Institute, Tokyo Institute of Technology,
2-12-1 Ookayama, Meguro, Tokyo, 152-8550, Japan
}

\altaffiltext{4}{
Department of Physics, The University of Tokyo, Tokyo, 113--0033, Japan
}

\altaffiltext{5}{
Department of Physics, and Kavli Institute for Astrophysics
and Space Research,\\
Massachusetts Institute of Technology, Cambridge, MA 02139, USA
}
\email{norio.narita@nao.ac.jp}


\maketitle

\begin{abstract}
We report new radial velocities of the TrES-4 transiting planetary system,
including observations of a full transit,
with the High Dispersion Spectrograph of the Subaru 8.2m telescope.
Modeling of the Rossiter-McLaughlin effect indicates that TrES-4b has
closely aligned orbital and stellar spin axes, with
$\lambda = 6.3^{\circ} \pm 4.7^{\circ}$.
The close spin-orbit alignment angle of TrES-4b
seems to argue against a migration history
involving planet-planet scattering or Kozai cycles,
although there are two nearby faint stars that could be
binary companion candidates.
Comparison of our out-of-transit data from 4 different runs
suggest that the star exhibits radial velocity variability of
$\sim$20~m~s$^{-1}$ in excess of a single Keplerian orbit.
Although the cause of the excess radial velocity variability is
unknown, we discuss various possibilities including systematic
measurement errors, starspots or other intrinsic
motions, and additional companions besides the transiting planet.
\end{abstract}
\footnotetext[*]{Based on data collected at Subaru Telescope,
which is operated by the National Astronomical Observatory of Japan.}

\section{Introduction}

Transiting planets provide us with valuable opportunities
to characterize exoplanetary systems.
One such opportunity is to measure the Rossiter-McLaughlin effect
(hereafter the RM effect: \cite{1924ApJ....60...15R},
\cite{1924ApJ....60...22M}), an apparent radial velocity anomaly
during a planetary transit, which is caused by a partial
eclipse of the rotating surface of the host star.
By measuring and modeling this effect, one can measure
the sky-projected angle
between the stellar spin axis and the planetary orbital axis.
Many theoretical investigations of the RM effect
have been presented
(e.g., \cite{2005ApJ...622.1118O, 2006ApJ...650..408G, 2007ApJ...655..550G,
2010ApJ...709..458H}),
and observations of the RM effect have been reported for
about 20 transiting planetary systems
(for the most recent compilation, see \cite{2010arXiv1001.0416J}
and references therein).
One of the main theoretical motivations to observe the RM effect is that
the observed degree of spin-orbit alignment is thought to be connected
with the migration history of the transiting planet.

The most frequently discussed planetary migration mechanisms are
(1) gravitational interaction between a protoplanetary disk
and a growing planet (disk-planet interaction models, e.g.,
\cite{1985prpl.conf..981L, 1996Natur.380..606L, 2004ApJ...616..567I}),
(2) gravitational planet-planet scattering and subsequent
tidal evolution
(planet-planet scattering models,
e.g., \cite{1996Sci...274..954R, 2002Icar..156..570M, 
2008ApJ...678..498N, 2008ApJ...686..580C}),  or
(3) the Kozai mechanism caused by a distant massive companion
and subsequent tidal evolution
(Kozai migration models,
e.g., \cite{2003ApJ...589..605W, 2005ApJ...627.1001T, 
2007ApJ...669.1298F, 2007ApJ...670..820W}).
These scenarios are not necessarily mutually exclusive, but to the extent
that they are, disk-planet interaction models would
predict small orbital eccentricities
and good spin-orbit alignments, while
planet-planet scattering models and Kozai migration models
predict a broader range of eccentricities and spin-orbit alignment angles.
Until about a year ago, all of the measurements indicated close alignments,
but recently 6 transiting planets have been reported with
significant misalignments:
XO-3b \citep{2008A&A...488..763H, 2009ApJ...700..302W},
HD~80606b \citep{2009A&A...498L...5M,
2009A&A...502..695P, 2009ApJ...703.2091W},
WASP-14b \citep{2009PASP..121.1104J, 2009MNRAS.392.1532J},
HAT-P-7b \citep{2009PASJ...61L..35N, 2009ApJ...703L..99W},
CoRoT-1b \citep{2009MNRAS.tmpL.360P}, and
WASP-17b \citep{2010ApJ...709..159A}.
With this increase in the number of measurements, and the
diversity of results, we are approaching the time when we may
test the validity and applicability of
the different planetary migration models.

The main target of this paper is TrES-4b,
which is a transiting exoplanet discovered by
\citet{2007ApJ...667L.195M} (hereafter M07) in the course of the TrES survey,
supplemented by Keck radial velocity measurements.
The planet orbits an F8 \citep{2009A&A...498..567D} host star
every 3.55 days and is one of the most ``inflated'' hot Jupiters
with a radius of about 1.8 $R_{\rm Jup}$,
which places this planet to be one of the least density exoplanets
ever discovered.
Refined spectroscopic and photometric characteristics of the host star TrES-4
were presented by \citet{2009ApJ...691.1145S} (hereafter S09).
The amplitude of the RM effect for TrES-4b was expected to be large,
because of the large projected equatorial rotational velocity
of the TrES-4 star ($V \sin I_s = 8.5$~km~s$^{-1}$; S09).
In addition, \citet{2009A&A...498..567D} have recently reported
a possible companion star around the TrES-4 system.
Although it has not yet been confirmed that the companion
is a true physical binary as opposed to a chance alignment,
a companion star would raise the possibility of migration via
Kozai cycles, lending additional motivation to the study
of the RM effect in this system.
We note that we adopt ``TrES-4'' as the host star name
and ``TrES-4b'' as the planet name in this paper,
although the planet was originally named ``TrES-4''
by the discoverers (see M07).
The reason of our choice is because recent papers on this system
(e.g., Daemgen et al. 2009, which we referred in this paper)
often describe the host star as ``TrES-4''.
Thus we consider that it would be confusing for readers
if we describe the planet as ``TrES-4'' in our paper,
and we hope to avoid such confusions.

In this paper, we present new measurements of the radial velocity
of TrES-4 made with the Subaru 8.2m telescope.
Although TrES-4 is relatively faint ($V = 11.6$),
the large aperture of the Subaru telescope has enabled us
to measure radial velocities of TrES-4 with high precision.
Our radial velocity dataset consists of 23 samples
covering a full transit, and 8 samples obtained outside of transits
on 3 different nights.
In addition to reporting the spin-orbit alignment angle of TrES-4b,
we also report the observation of radial velocity variation in excess
of the previously observed single Keplerian orbit,
and we confirm through direct imaging with the HDS slit viewer that
there are candidate companion stars.

The rest of this paper is organized as follows.
Section~2 summarizes our Subaru observations, and
section~3 describes analysis procedures of the RM effect of TrES-4b.
Section~4 presents our main result on the spin-orbit alignment angle
of TrES-4b, and
section~5 discusses possible causes of the additional radial velocity
variation in this system.
Finally, section~6 summarizes the findings of this paper.

\section{Observations}

We observed a full transit of TrES-4b with the High Dispersion
Spectrograph (HDS: \cite{2002PASJ...54..855N}) on the Subaru 8.2m
telescope on UT 2007 July 13.
In addition, we measured radial velocities
on UT 2007 August 5, UT 2008 March 9, and UT 2008 May 30,
when transits were not occurring.
For all observations, we employed the same setup.
We used the standard I2a setup of the HDS,
a slit width of $0\farcs8$ corresponding to
a spectral resolution of about 45,000, and
the iodine gas absorption cell for precise differential
radial velocity measurements.
The exposure times for the radial velocity measurements
were 12--30 minutes, yielding a typical signal-to-noise ratio
(SNR) of approximately 80--100 per pixel.
We processed the observed frames with standard IRAF\footnote{The Image
  Reduction and Analysis Facility (IRAF) is distributed by the U.S.\
  National Optical Astronomy Observatories, which are operated by the
  Association of Universities for Research in Astronomy, Inc., under
  cooperative agreement with the National Science Foundation.}
procedures and extracted one-dimensional spectra.
We note that in about half of the exposures,
light from a nearby companion star (see section 3.1)
was also admitted through the slit.
For this reason, when extracting the spectra we limited
the aperture width of echelle orders
to 3--4 pixels, to avoid including any significant flux
from the companion star.
We computed relative radial velocities and uncertainties
following the algorithm of
\citet{1996PASP..108..500B} and \citet{2002PASJ...54..873S},
as described in \citet{2007PASJ...59..763N}.
We estimated the internal measurement uncertainty of each radial velocity
based on the scatter of radial velocity solutions
for $\sim$4~\AA~segments of each spectrum.
The typical internal uncertainties were 15--20~m~s$^{-1}$,
which is worse than typical cases having similar SNR
due to relatively rapid rotation of TrES-4.
The radial velocities and uncertainties are summarized in Table~1.

\section{Analyses}

\subsection{HDS Slit Viewer Images}

The slit viewer of the Subaru HDS has a 512$\times$512 CCD,
providing unfiltered 60''$\times$60'' field of view images.
During the HDS observations, we found two nearby companion stars
in HDS slit viewer images.
One is the companion star
which \citet{2009A&A...498..567D} also reported
to the north of TrES-4, and the other is a newly-discovered star
located in west-southwest.
Figure~1 shows 4 magnified portions of slit viewer images obtained
on UT 2007 July 13 (upper left), UT 2007 August 5 (lower left),
UT 2008 March 9 (upper right), and UT 2009 July 12 (lower right).
North is up and east is left for these images, and the field of view
is 20''$\times$20''.
Our ability to perform astrometry and photometry on these images
is limited, since TrES-4 located on the slit.
For the north companion star,
we roughly estimate the separation angle as $\sim 1\farcs56$
($\sim13$ pixels) and the position angle as $\sim 0^{\circ}$
(almost true north).
These findings are consistent
with those of \citet{2009A&A...498..567D}.
While the west-southwest companion star is located at
the separation angle of $\sim 7\farcs8$
($\sim65$ pixels) and the position angle of $\sim 249^{\circ}$.
We also estimate that the companion stars are at least
4 magnitudes fainter than TrES-4 in visible wavelength.
Although neither our data nor the data of \citet{2009A&A...498..567D}
can be used to tell whether or not the companion star is
physically associated with TrES-4, it may be possible
to do so with multiband IR (for example, \textit{JHK} band)
photometry with adaptive optics, which
would be very useful to study common proper motion,
color-magnitude relation, and spectral type of
the companion stars.
Such follow-up observations will enable us to
investigate the binarity of the companion stars.

\subsection{Simulated Formula for the Rossiter-McLaughlin Effect}

We modeled the RM effect of TrES-4 following the
procedure of \citet{2005ApJ...631.1215W}, as adapted for HDS by
Narita~et~al.~(2009a, 2009b)
and discussed further by \citet{2010ApJ...709..458H}.
We first created a synthetic template spectrum which matches
the stellar properties of TrES-4 described by S09,
using a model by \citet{2005A&A...443..735C}.
To simulate the disk-integrated spectrum of TrES-4,
we applied a rotational broadening kernel of $V \sin I_s =
8.5$~km~s$^{-1}$ (S09) and adopted the quadratic limb-darkening
parameters for the spectroscopic band as $u_1 = 0.46$ and $u_2 = 0.31$
based on the tables of \citet{2004A&A...428.1001C}.
To simulate in-transit spectra of TrES-4,
we subtracted a scaled-down and velocity-shifted copy of
the original unbroadened
spectrum, which represents the hidden part of the stellar surface.
We created a collection of such simulated spectra
for various values of the scaling factor $f$ and the
velocity-shift $v_p$, and computed the apparent radial velocity
$\Delta v$ for each spectrum.
We fitted $\Delta v$ in ($f, v_p$) space and
determined an empirical formula of the RM effect for TrES-4 as
\begin{equation}
\Delta v = - f v_p \left[1.623 - 0.885
\left( \frac{v_p}{V \sin I_s} \right)^2 \right].
\end{equation}

\subsection{Radial Velocity Modeling}

Since we did not have good transit light curves for TrES-4b,
we fixed stellar and planetary parameters of TrES-4 to
the values reported by S09 as follows;
the stellar mass $M_s = 1.404$ [$M_{\odot}$],
the stellar radius $R_s = 1.846$ [$R_{\odot}$],
the radius ratio $R_p/R_s = 0.09921$,
the orbital inclination $i = 82.59^{\circ}$,
and the semi-major axis in units of the stellar radius
$a / R_s = 5.94$.
As reported in Narita~et~al.~(2009a, 2009b),
these assumptions might lead a certain level of systematic errors
in results due to uncertainties in the fixed parameters,
especially in $i$ and $a / R_s$.
We estimated such systematic errors in section~4.
We also fixed the transit ephemeris $T_c = 2454230.9053$ [HJD]
and the orbital period $P = 3.553945$~days based on S09.
Although this ephemeris had an uncertainty of 3 minutes for the
observed transit, the uncertainty was well within our time-resolution
(exposure time of 12--30 minutes and readout time of 1 minute)
and thus negligible.
The adopted parameters above are summarized in table~2.

Our model had 3 free parameters describing the TrES-4 system:
the radial velocity semiamplitude ($K$),
the sky-projected stellar rotational velocity ($V \sin I_s$),
and the sky-projected angle between the stellar spin axis and
the planetary orbital axis ($\lambda$).
We also added two offset velocity parameters for respective radial
velocity datasets ($v_1$: our Subaru dataset, $v_2$: Keck in M07).
Note that we fixed the eccentricity ($e$) to zero at first,
and the argument of periastron ($\omega$) was not considered.
The assumption of zero eccentricity is reasonable
since \citet{2009ApJ...691..866K} constrained 
$e \cos \omega < 0.0058$ ($3\sigma$)
based on Spitzer observations of the secondary eclipse.

We then calculated the $\chi^2$ statistic
\begin{eqnarray}
\chi^2 &=& \sum_i \left[ \frac{v_{i,{\rm obs}}-v_{i,{\rm calc}}}
{\sigma_{i}} \right]^2,
\end{eqnarray}
where
$v_{i, {\rm obs}}$ were the observed radial velocity data
and $v_{i, {\rm calc}}$ were the values calculated based on
a Keplerian motion and on the RM formula given above.
$\sigma_{i}$ were calculated by the quadrature sum of
the internal errors of the observed radial velocities
and expected stellar jitter level of $4.4$~m~s$^{-1}$.
We note that we used the mean value of the case of
$\Delta M_V < 1, \Delta F_{{\tiny{\textrm{Ca} \textsc{II}}}} < 0.6,
B - V < 0.6$ in Wright (2005) for the jitter,
based on the stellar properties reported by S09.
We determined optimal orbital parameters by minimizing the $\chi^2$
statistic using the AMOEBA algorithm \citep{1992nrca.book.....P}.
We estimated 1$\sigma$ uncertainty of each free parameter based on
the criterion $\Delta \chi^2 = 1.0$.

\section{Results}

We first fitted all Subaru radial velocity samples
with the published 4 radial velocities presented in M07.
The upper panels of figure~2 show the radial velocities
as a function of orbital phase plotted with the best-fit model curve,
and the lower panels plot the radial velocities as a function of HJD.
Best-fit parameter values, uncertainties, and the reduced $\chi^2$
for this fit are summarized in the left column of table~3.
The fit is poor, with $\chi^2=47.4$ and 30 degrees
of freedom ($\chi^2_\nu = 1.58$).
Apparently there is an inconsistency between
the Subaru out-of-transit velocities, and the M07 Keck velocities,
even after allowing for a constant offset between these data sets.
Comparison of all the observations suggests that there
are additional out-of-transit radial velocity variations
of about 20~m~s$^{-1}$, which is the root-mean-squared (rms) residual
of the Subaru and Keck data.
Although the reason for this excess variability is still unclear,
some possible reasons are long-term instrumental instabilities,
starspots, intrinsic motions of the stellar surface (``stellar jitter''),
and the presence of other bodies in the TrES-4 system.
We discuss these possibilities in section 5. Here,
we concentrate on a reasonable model of the RM effect
in spite of the excess variability, to
give our best estimate for the sky-projected spin-orbit angle of
the transiting planet TrES-4b.

Our approach was to use the Subaru data from
the transit night (UT~2007 July 13),
along with the Keck data of M07. We chose not to fit the Subaru data
obtained on the other 3 nights outside of transits.
The left panel of figure~3 plots the data and the best-fit curve
(the solid line), and the right panel shows a zoom of the RM effect.
The results are summarized in the middle column of table~3.
In this case, the fit is good ($\chi^2_\nu = 0.45$) and
the best-fit model indicates a good spin-orbit alignment,
with $\lambda = 7.3^{\circ} \pm 4.6^{\circ}$.
The best-fit model also gives
$V \sin I_s = 8.3 \pm 1.1$~km~s$^{-1}$, in agreement with
the S09 result based on a spectroscopic line analysis.
We note that some of the early RM analyses using
the analytic formula for the RM effect
(e.g., \cite{2005ApJ...622.1118O,2006ApJ...650..408G})
tended to overestimate the stellar rotational velocity
$V \sin I_s$ (e.g., \cite{2005ApJ...631.1215W}).
Recently \citet{2010ApJ...709..458H} and \citet{2009arXiv0911.5361C}
studied the reason for the discrepancy, and \citet{2010ApJ...709..458H}
reported an improved method to address this problem.
The agreement between our RM results and the spectroscopic analysis of S09
suggests that the RM calibration process explained above 
works well.
As for the radial velocity semiamplitude, the fit indicates
$K = 94.9 \pm 7.2$~m~s$^{-1}$
(the M07 result was $K = 97.4 \pm 7.2$~m~s$^{-1}$ for reference).

In addition, we estimated the sensitivity of our results to the choices
of the fixed photometric parameters by fitting the radial velocities
using other choices for those parameters:
(1) $a / R_s = 6.15,\,\, i = 82.99^{\circ}$
(corresponding to 1$\sigma$ lower limit of the impact parameter in S09);
(2) $a / R_s = 5.73,\,\, i = 82.19^{\circ}$
(corresponding to 1$\sigma$ upper limit of the impact parameter in S09).
Consequently, we found that respective results for
$\lambda$ and $V \sin I_s$ are;
(1) $\lambda = 7.2^{\circ} \pm 4.5^{\circ}$
and $V \sin I_s = 8.6 \pm 1.1$~km~s$^{-1}$;
(2) $\lambda = 7.5^{\circ} \pm 4.8^{\circ}$
and $V \sin I_s = 8.0 \pm 1.1$~km~s$^{-1}$.
Thus the conclusion of the spin-orbit alignment does not change,
and the uncertainties in the photometric parameters have very little
effects on our results.

We noticed that the residuals of the Subaru dataset have a small
positive gradient with time (see the right middle panel of figure~3).
This indicates that the Subaru data alone prefer
a slightly smaller value of $K$
than the joint fit of Subaru and Keck data.
We therefore fitted the radial velocities of
the Subaru UT 2007 July 13 dataset only, for reference.
The best-fit model is shown by the dotted lines in figure~3,
and derived parameters are summarized in the right column of table~3.
The data indicate a smaller radial velocity
semiamplitude, but with a larger uncertainty: $K = 64.6 \pm 27.7$~m~s$^{-1}$.
This result may be supporting evidence that a true $K$ for the TrES-4b
is actually smaller, although it is not very convincing.
In this light, it is interesting to point out that TrES-4b
was previously known as the lowest density planet
($\rho = 0.202^{+0.038}_{-0.032}$~g~cm$^{-3}$: S09).
Recently WASP-17b ($\rho = 0.092^{+0.054}_{-0.032}$~g~cm$^{-3}$:
\cite{2010ApJ...709..159A})
and Kepler-7b ($\rho = 0.166^{+0.019}_{-0.020}$~g~cm$^{-3}$:
\cite{2010arXiv1001.0190L})
were reported to have lower densities than TrES-4b,
and thus TrES-4b is currently the third lowest density planet
discovered so far.
The observed out-of-transit radial velocity variation and
the radial velocity gradient around the transit phase
may suggest a lower density of TrES-4b than previously reported.
Thus it is important to obtain radial velocities of TrES-4
not only to confirm the additional radial velocity variation,
but also to measure the density of TrES-4b more precisely.
We note that this fit gives a small reduced chi-squared
($\chi^2_\nu = 0.35$) and $\lambda = 5.3^{\circ} \pm 4.7^{\circ}$,
indicating a good spin-orbit alignment as before.

We adopt a compromise value of the latter 2 models as our
representative result, namely $\lambda = 6.3^{\circ} \pm 4.7^{\circ}$.
The small difference between the results of the latter 2 models
shows that there is a small systematic uncertainty due
to the choice of model.
Consequently, we conclude that the TrES-4b
has a small value of $\lambda$, based on the
model of the RM effect.
However, since we could not find a satisfactory
solution that explains the all observed radial velocities
(including all of the Subaru data and the M07 data),
further radial velocity measurements are desired
to understand any excess variability and to give
greater confidence to the orbital solution.

\section{Discussion}

\subsection{Possible Causes of the Additional Radial Velocity Variation}

Since we could not find an appropriate model for the all observed
radial velocities at this point (see figure~2),
we here discuss possibilities of systematic effects
as well as real sources of excess radial velocity variation.

\subsection*{-- Instrumental Instability of the Subaru HDS}

Since the Subaru radial velocities were gathered on a few days
in clusters spanning about 1 year,
it is of utmost importance to know the instrumental stability of the HDS.
For observations within a single night,
\citet{2007PASJ...59..763N} studied the radial velocity standard star
HD~185144 and found that the Subaru HDS is stable within a few~m~s$^{-1}$.
In addition, \citet{2009ApJ...703L..99W} reported that
the Subaru HDS is stable within a few~m~s$^{-1}$ over two weeks
based on HAT-P-7 observations. Likewise, \citet{2008ApJ...686..649J} 
did not find systematic offsets for the HAT-P-1 system over approximately
1 month, using the same setup (and even some of the same nights)
as the TrES-4 observations presented here.
However, specific stability over 1 year has not yet been
confirmed through monitoring of radial velocity standard stars,
although studies for such long-term stability of
the Subaru HDS are in progress.
Thus we note that the instrumental instability of the Subaru HDS
is one of the prime possibilities of a cause of the additional
radial velocity variation at this point.

\subsection*{-- Starspots}

One possibility involves starspots on the photosphere of TrES-4.
Since the rotational velocity of TrES-4 is relatively fast ($V \sin
I_s = 8.5$~km~s$^{-1}$: S09), stellar spots of similar size to a
planet would cause an apparent radial velocity shift like the RM
effect, on a timescale of the stellar rotation period ($P_{\rm rot}
\approx 11$~days, assuming $\sin I_s \approx 1$).
For example, a dark spot of approximately the same size as the planet
would lead to a maximum shift of 85~m~s$^{-1}$, while smaller spots
with less contrast would contribute smaller velocities.
If this is the case, all the RV data are affected.
However, one would not expect a hot F8 star to have large spots.
It is because M07 did not report such possibility of stellar spots
from the TrES transit survey,
S09 reported no active $\textrm{Ca}$ HK line emission
($\log R'_{\tiny{\textrm{HK}}} = -5.11 \pm 0.15$),
and \citet{2009ApJ...691..866K} concluded
that spot activity is unlikely (but not impossible) based on
Spitzer observations.
Thus the spot explanation is doubtful, although
further long-term photometric monitoring would be useful to
constrain this hypothesis still further.

\subsection*{-- Other Sources of Stellar Jitter}

Another possible explanation of the systematic radial velocity variation
is an intrinsic stellar jitter (see e.g., \cite{2005PASP..117..657W}), i.e.,
motions of the stellar photosphere due to pulsations or other flows.
Although an empirical relation reported by \citet{2005PASP..117..657W}
predicted a typical stellar jitter of $4.4$~m~s$^{-1}$ for stars like TrES-4,
it is conceivable that TrES-4 has an unusually unstable photosphere.
To explain the observed radial velocities, we would need to
invoke a stellar jitter of about $20$~m~s$^{-1}$ for TrES-4
based on the rms residuals of the Subaru and Keck datasets.
The jitter would need to have a time scale longer than a few days,
in order to explain why the M07 observations (conducted on
consecutive three nights) do not exhibit such a large scatter.
In this light the hypothesis that the stellar jitter explains all
the excess variability seems too contrived.

\subsection*{-- Contamination of Companion Star's Lights or Sky Backgrounds}

As described by \citet{1996PASP..108..500B}, a slight change in
the instrumental profile would result in a systematic shift of
the apparent radial velocity.
Thus, any contaminating light from the nearby candidate
companion star of TrES-4 or sky backgrounds (e.g., moonlight or twilight)
may have affected the radial velocities.
We limited the aperture width
of echelle orders in order to avoid contamination of
lights from the north companion star.
Moonlight contamination is unlikely
since our TrES-4 observations were conducted in clear and moonless time,
and sky background levels were still low
although 2 exposures (HJD of 2454317.74338 and 2454535.14401)
were conducted during twilight.
Note that M07 did not report the existence of the companion star,
and therefore it is possible that the companion star might
have been on the slit during the M07 observations.
Such an effect might have caused some systematic shifts in the M07 data.
Although we were not able to estimate the systematic effect in the M07 data,
it could be a small effect since the companion is very faint.

\subsection*{-- Eccentricity of TrES-4b}

As M07 reported only 4 radial velocity samples, they did not include
the eccentricity $e$ and the argument of periastron $\omega$
in their radial velocity model. Instead they assumed the orbit
to be circular as we have done, and more recently
\citet{2009ApJ...691..866K} found
$e \cos \omega < 0.0058$ ($3\sigma$)
based on Spitzer secondary eclipse observations.
With the Subaru data we now have a sufficient number of radial velocity samples
to allow the eccentricity and argument of pericenter to be free parameters.
However, allowing $e$ and $\omega$ to vary does not improve the model fit,
and the eccentricity of the best-fit model is nearly zero.
This is consistent with the constraint by \citet{2009ApJ...691..866K}.
Thus a large eccentricity of TrES-4b could not be the explanation for the
observed excess RV variability.

\subsection*{-- Additional Planets}

If the preceding explanations for the observed radial velocity variation
could be ruled out, we would consider a possibility of
presence of additional planets.
This is the most interesting case, however, at this point
our Subaru observations were too sparse to find and confirm
another periodicity in the radial velocities.
We can only conclude that the radial velocity semiamplitude of about
20~m~s$^{-1}$ over a year is possible for hypothetical planets
in the TrES-4 system.
Obviously, further continuous radial velocity monitoring
would be necessary to check on this possibility.

\section{Summary}

We observed radial velocities of TrES-4 including a full transit
of TrES-4b with the HDS of the Subaru 8.2m telescope.
Our radial velocity modeling has revealed that TrES-4b has
a close alignment between the projected orbital and stellar spin
axes, based on the RM effect.
On the other hand, we could not find a satisfactory Keplerian model
that agrees with all the data.
Although the true cause of the excess radial velocity variation is
still unclear, systematic errors in long-term Subaru observations,
stellar spots, a large stellar jitter, or additional planets might
play a role in the observed radial velocities.
The small spin-orbit alignment angle as well as the small eccentricity
of TrES-4b seems to match migration models considering
disk-planet interactions
rather than planet-planet scattering models or Kozai migration models.
Although TrES-4 has the binary candidates to the north
and the west-southwest,
we did not find supporting evidence of the Kozai migration.
To confirm and discriminate the true cause of the radial velocity
variation of TrES-4, further radial velocity monitoring and photometric
monitoring are highly desired.

\bigskip
We acknowledge the invaluable support for our Subaru observations
by Akito Tajitsu, a support scientist for the Subaru HDS.
We are deeply grateful for David Charbonneau and the TrES collaboration,
who kindly informed us the transit ephemeris of TrES-4b
in advance of the publication of the M07 paper.
We appreciate a careful reading and insightful
comments by the anonymous referee.
This paper is based on data collected at Subaru Telescope,
which is operated by the National
Astronomical Observatory of Japan.  The data analysis was in part
carried out on common use data analysis computer system at the Astronomy
Data Center, ADC, of the National Astronomical Observatory of Japan.
N.N. is supported by a Japan Society for Promotion of Science (JSPS)
Fellowship for Research (PD: 20-8141), and was also supported in part
by the National Science Foundation under Grant No. NSF PHY05-51164
(KITP program ``The Theory and Observation of Exoplanets'' at UCSB).
We wish to acknowledge the very significant cultural role
and reverence that the summit of Mauna Kea has always had within
the indigenous Hawaiian community.



\clearpage

\begin{figure*}[pthb]
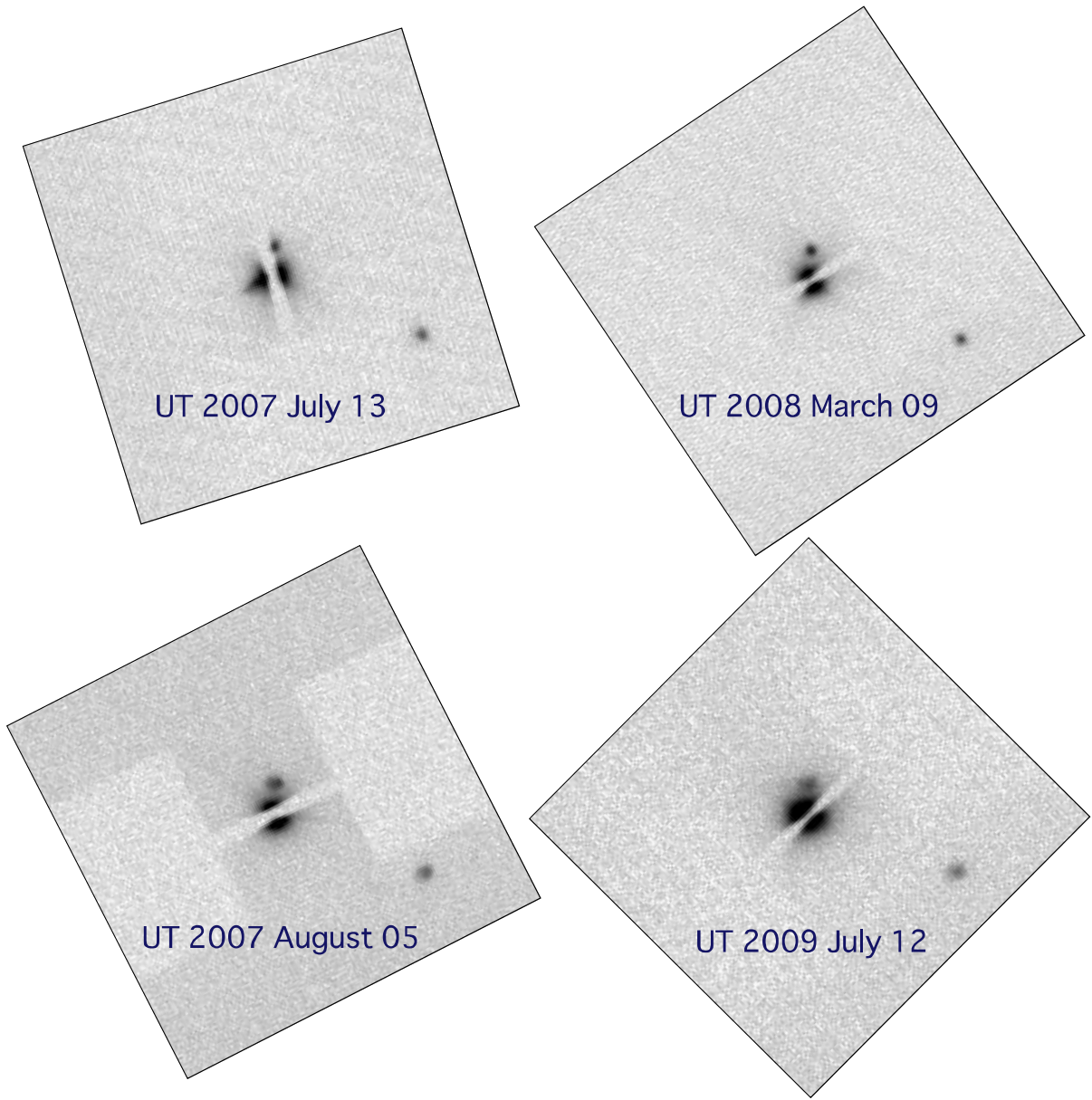

 \begin{center}
  \FigureFile(160mm,160mm){figure1.eps}
 \end{center}
  \caption{HDS slit viewer images of TrES-4 taken on
  UT 2007 July 13 (upper left), UT 2007 August 05 (lower left),
  UT 2008 March 9 (upper right), and UT 2009 July 12 (lower right).
  North is up and east is left, and field of view is $20"\times20"$
  for the all images.
  }
\end{figure*}

\begin{figure*}[pthb]
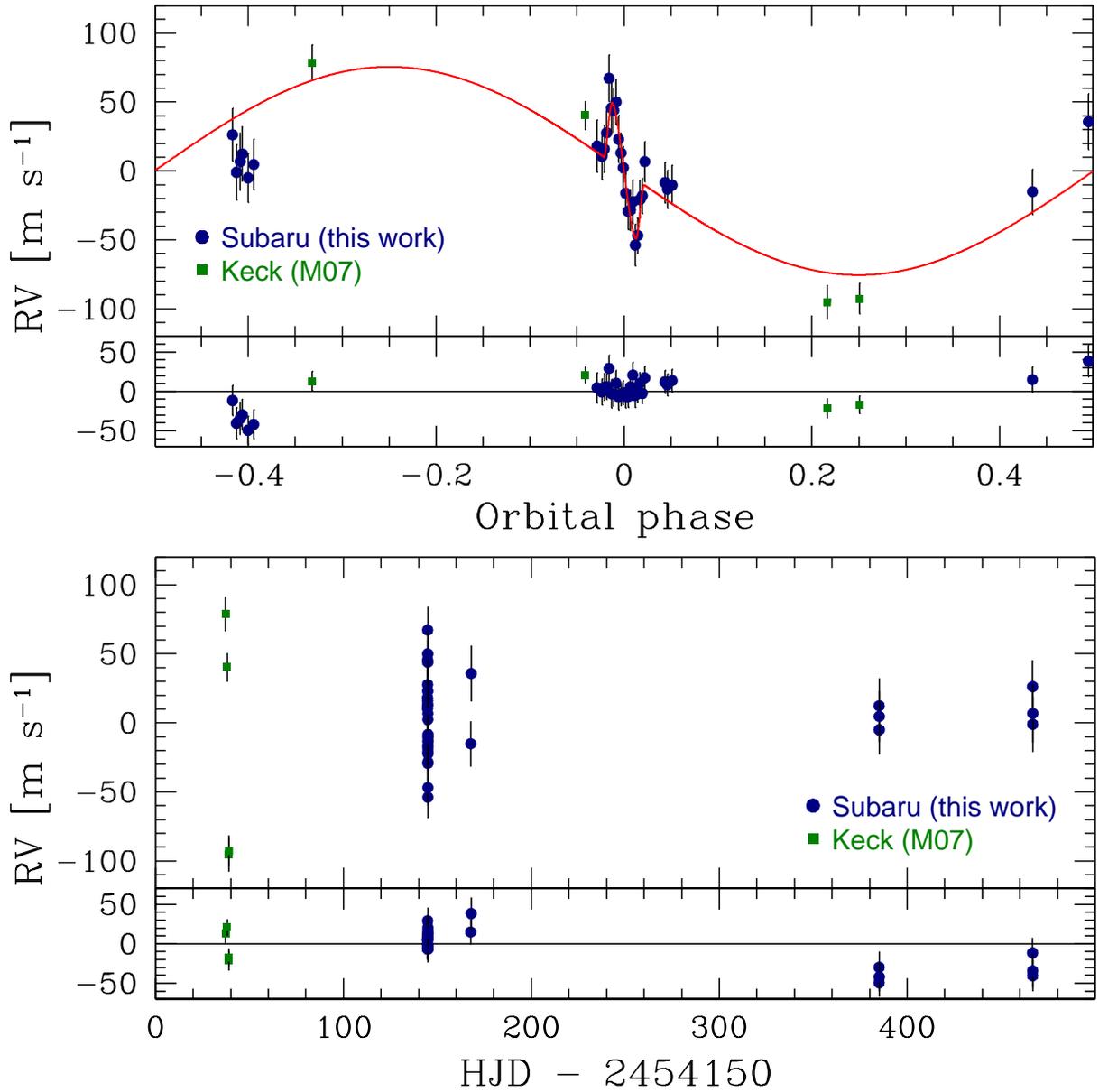

 \begin{center}
  \FigureFile(160mm,160mm){figure2a.eps}
  \FigureFile(160mm,160mm){figure2b.eps}
 \end{center}
  \caption{
  Radial velocities (RVs) and the best-fit curve of TrES-4
  as a function of orbital phase (upper) and as a function of HJD (lower).
  All Subaru RVs and the M07 RVs are used.
  Bottom panels: Residuals from the best-fit curve.
  }
\end{figure*}

\begin{figure*}[pthb]
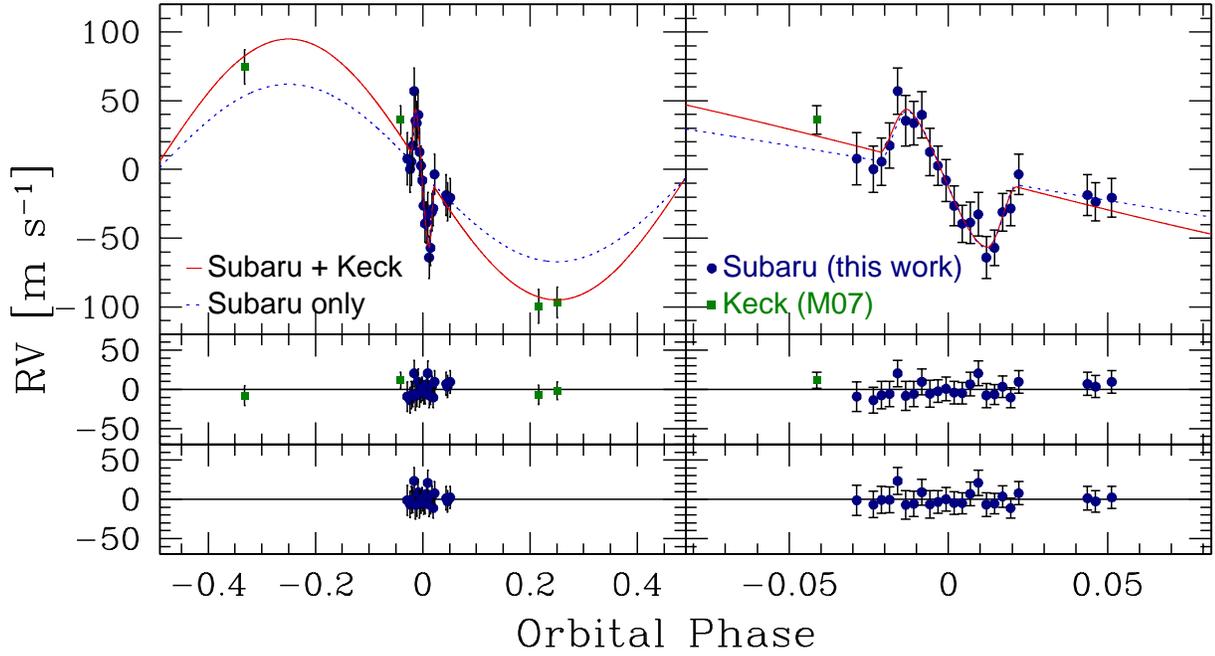

 \begin{center}
  \FigureFile(160mm,160mm){figure3.eps}
 \end{center}
  \caption{
  Top panels:
  Radial velocities (RVs) and best-fit curves of TrES-4
  as a function of orbital phase. The left panels show the entire orbit
  and the right panels are the zoom of transit phase.
  The solid line indicates the best-fit curve for the case that
  Subaru RVs taken on UT 2008 July 13 and the M08 RVs are used,
  and the dotted line is for the case that only Subaru RVs are used.
  Middle panels: Residuals from the solid model curve.
  Bottom panels: Residuals from the dotted model curve.
  }
\end{figure*}

\begin{table}[htb]
\caption{Radial velocities obtained with the Subaru/HDS.}
\begin{center}
\begin{tabular}{lcc}
\hline
Time [HJD]  & Value [m~s$^{-1}$] & Error [m~s$^{-1}$]\\
\hline
2454294.7715 & -5.1 & 18.6 \\
2454294.7901 & -12.7 & 16.2 \\
2454294.7991 & -7.3 & 16.7 \\
2454294.8081 &  4.5 & 15.8 \\
2454294.8171 & 44.1 & 16.3 \\
2454294.8261 & 22.5 & 17.8 \\
2454294.8351 & 20.7 & 15.4 \\
2454294.8441 & 26.8 & 16.1 \\
2454294.8531 & -0.2 & 16.6 \\
2454294.8621 & -10.2 & 13.7 \\
2454294.8711 & -20.9 & 14.6 \\
2454294.8801 & -39.3 & 13.8 \\
2454294.8891 & -52.5 & 12.8 \\
2454294.8981 & -51.6 & 14.5 \\
2454294.9071 & -45.5 & 15.2 \\
2454294.9161 & -77.1 & 14.6 \\
2454294.9252 & -70.0 & 12.2 \\
2454294.9342 & -43.9 & 13.0 \\
2454294.9432 & -41.3 & 11.9 \\
2454294.9522 & -16.5 & 14.0 \\
2454295.0289 & -31.6 & 14.2 \\
2454295.0379 & -36.5 & 13.1 \\
2454295.0559 & -33.5 & 13.6 \\
2454317.7434 & -38.4 & 15.8 \\
2454317.9553 & 12.6  & 19.6 \\
2454535.1010 & -10.9 & 19.3 \\
2454535.1225 & -28.2 & 17.3 \\
2454535.1440 & -18.4 & 18.0 \\
2454616.8015 &   3.0 & 18.8 \\
2454616.8162 & -24.2 & 19.6 \\
2454616.8309 & -16.4 & 20.3 \\
\hline
\end{tabular}
\end{center}
\end{table}

\begin{table}[t]
\caption{Adopted stellar and planetary parameters.}
\begin{center}
\begin{tabular}{l|cc}
\hline
Parameter & Value & Source \\
\hline
$M_s$ [$M_{\odot}$] 
& $1.404$ & S09 \\
$R_s$ [$R_{\odot}$]
& $1.846$ & S09 \\
$R_p/R_s$ 
& $0.09921$ & S09 \\
$i$ [$^{\circ}$]
& $82.59$ & S09 \\
$a / R_s$
& $5.94$ & S09 \\
$u_1$
& $0.46$ & \citet{2004A&A...428.1001C} \\
$u_2$
& $0.31$ & \citet{2004A&A...428.1001C} \\
jitter [m~s$^{-1}$]
& $4.4$ &  \citet{2005PASP..117..657W} \\
$T_c$ [HJD]
& $2454230.9053$ & S09 \\
$P$ [days]
& $3.553945$ & S09 \\
\hline
\multicolumn{3}{l}{\hbox to 0pt{\parbox{80mm}{\footnotesize
}\hss}}
\end{tabular}
\end{center}
\end{table}

\begin{table*}[t]
\caption{Best-fit values and uncertainties of the free parameters.}
\begin{center}
\begin{tabular}{l|cc|cc|cc}
\hline
 & \multicolumn{2}{c|}{Subaru all + Keck}
 & \multicolumn{2}{c|}{Subaru transit + Keck}
 & \multicolumn{2}{c}{Subaru transit only} \\
Parameter & Value & Uncertainty & Value & Uncertainty & Value & Uncertainty \\
\hline
$K$ [m s$^{-1}$] 
& 75.5 & $\pm 6.3$ 
& 94.9 & $\pm 7.2$ 
& 64.6 & $\pm 27.7$ \\
$V \sin I_s$ [km s$^{-1}$]
& 8.3  & $\pm 1.1$
& 8.3  & $\pm 1.1$
& 8.7  & $\pm 1.2$ \\
$\lambda$ [$^{\circ}$]
& 0.0  & $\pm 4.2$
& 7.3  & $\pm 4.6$
& 5.3  & $\pm 4.7$ \\
$v_1$ (Subaru) [m s$^{-1}$] 
& -23.2  & $\pm 3.8$
& -13.0  & $\pm 4.6$
& -15.5  & $\pm 5.1$\\
rms (Subaru) [m s$^{-1}$] 
& 19.9  & --
& 9.2  & --
& 8.4  & --\\
$v_2$ (Keck) [m s$^{-1}$] 
& 18.7  & $\pm 5.9$
& 22.9  & $\pm 6.0$
& --  & --\\
rms (Keck) [m s$^{-1}$] 
& 18.4  & --
& 7.9  & --
& --  & --\\
$\chi^2/\nu$ 
& 47.36/30  & --
& 9.88/22  & --
& 6.64/19  & --\\
\hline
\multicolumn{5}{l}{\hbox to 0pt{\parbox{100mm}{
}\hss}}
\end{tabular}
\end{center}
\end{table*}

\end{document}